# The form-invariance of wave equations without requiring a priori relations between field variables

By Zhihai Xiang*

*AML, Department of Engineering Mechanics, Tsinghua University, Beijing* 100084, *China*

According to *the principle of relativity*, the equations describing the laws of physics should have the same forms in all admissible frames of reference, i.e., form-invariance is an intrinsic property of correct wave equations. However, so far in the design of metamaterials by transformation methods, the form-invariance is always proved by using certain relations between field variables before and after coordinate transformation. The main contribution of this paper is to give general proofs of form-invariance of electromagnetic, sound and elastic wave equations in the global Cartesian coordinate system without using any assumption of the relation between field variables. The results show that electromagnetic wave equations and sound wave equations are intrinsically form-invariant, but traditional elastodynamic equations are not. As a by-product, one can naturally obtain new elastodynamic equations in the time domain that are locally accurate to describe the elastic wave propagation in inhomogeneous media. The validity of these new equations is demonstrated by some numerical simulations of a perfect elastic wave rotator and an approximate elastic wave cloak. These findings are important for solving inverse scattering problems in many fields such as seismology, nondestructive evaluation and metamaterials.

## 1. Introduction

Recently, exciting progress has been made on controlling the wave propagation with meta-devices, such as cloaks, rotators, benders, super-lenses, etc [1-3]. These devices are made from engineered metamaterials with the special distribution of refractive indices, so that they can steer the wave along the desired trajectory. To determine the effective material properties of these meta-devices, one has to solve an inverse scattering problem, which may have non-unique solutions [4, 5]. A novel idea to find a possible solution to this inverse problem is using the coordinate transformation method or change of variables [1-9], which was firstly proposed by Ward & Pendry [10] to solve Maxwell equations in complex geometries. With this method, wave equations in a virtual space $\Omega$ are transformed into a physical space $\Omega'$ according to a coordinate mapping $\boldsymbol{x}'=\boldsymbol{x}'(\boldsymbol{x})$ (see Figure 1) and specified relations between field variables in both spaces. If the forms of these equations do not change in the global Cartesian coordinate system, the effective material properties can be obtained by comparing the corresponding terms in the original and transformed equations. In this way, even the material is homogeneous and isotropic in the virtual space; the obtained material in the physical space is inhomogeneous and usually heterogeneous. In a word, if a wave equation is locally accurate for inhomogeneous and heterogeneous media, it should be form-invariant after arbitrary coordinate transformation in the global Cartesian coordinate system and can be used to design perfect metamaterials. This also complies with *the principle of material frame-indifference* [11].

* Tel: +86-10-62796873; *Email address*: xiangzhihai@tsinghua.edu.cn.

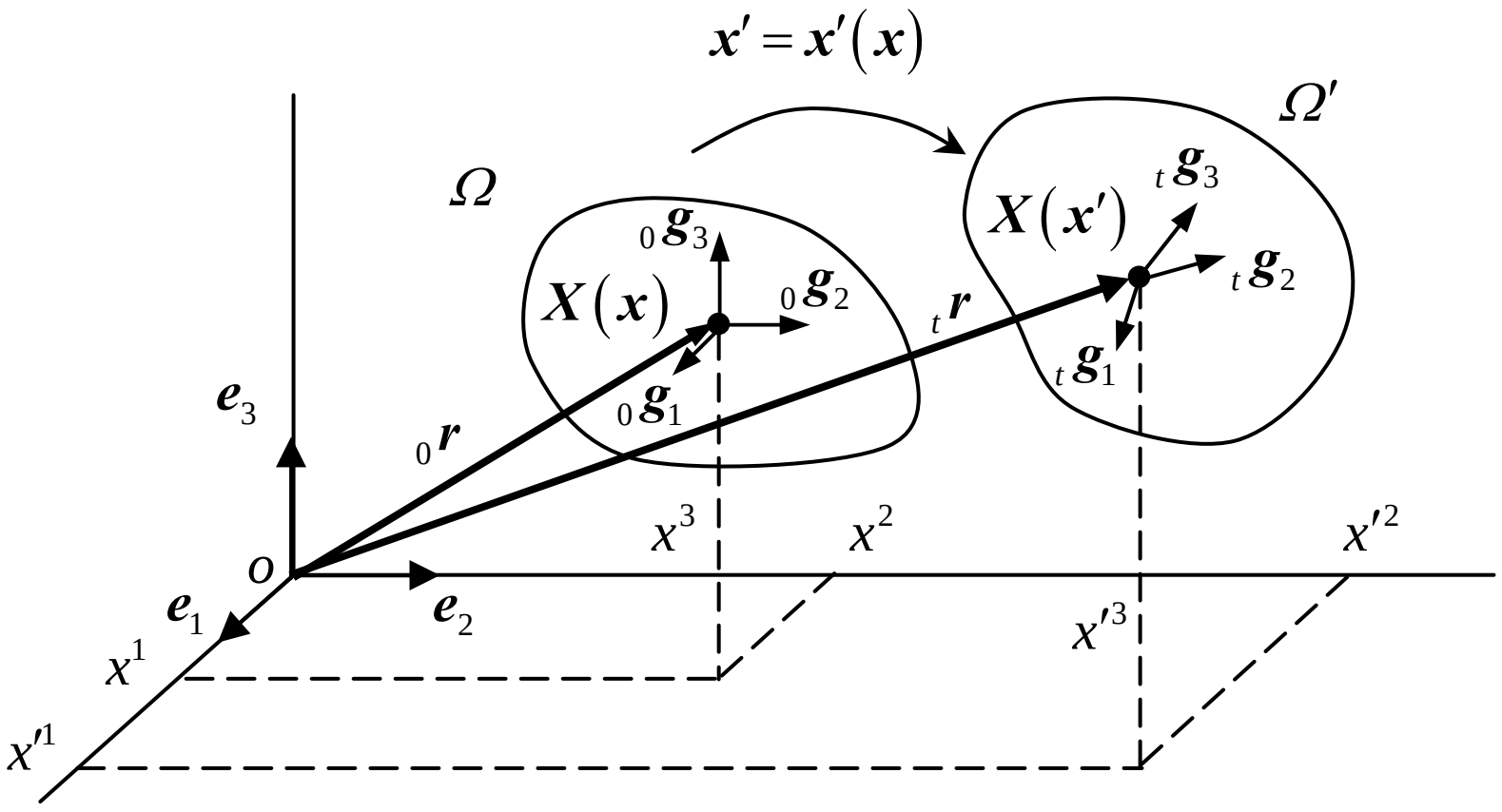

**Figure 1** Schematic of the transformation method.

It has been proved that Maxwell equations are form-invariant under arbitrary coordinate transformation [2, 6]. The sound equations are also form-invariant by their analogies with Maxwell equations [7, 8, 12]. However, the form-invariance is not generally valid for the elastodynamic equations [9]. It is also noticed that these proofs of form-invariance use deformation gradient to connect field variables before and after coordinate transformation. In this way, the deformed grid in physical space, which is Cartesian in virtual space, shows the wave trajectories [6]. By relaxing this constraint and assuming arbitrary linear gauge between field variables in the virtual and physical spaces, one can have more freedom to design metamaterials [13, 14]. However, according to *the principle of relativity*, the equations describing the laws of physics should have the same forms in all admissible frames of reference [15]. I.e., form-invariance is an intrinsic property of wave equations, which should be independent of the relations between field variables before and after coordinate transformation.

Since the early 1980's, Willis realized that traditional wave equations could not be accurate for inhomogeneous media. He establishes the theories of wave propagation in random media, based on variational principle [16-18]. The wave equations in Willis form contain additional non-local operators in contrast to traditional wave equations and can describe the non-local behavior of wave propagation in inhomogeneous media [19-21]. And recently, it was proved that Willis elastic wave equations are form-invariant and could have local approximations in the frequency domain [9]

$$\boldsymbol{\Sigma} = \boldsymbol{C} : \nabla \boldsymbol{U} + \boldsymbol{S} \cdot \boldsymbol{U} \tag{1}$$

$$\nabla \cdot \boldsymbol{\Sigma} = \boldsymbol{S}^{\mathrm{T}} : \nabla \boldsymbol{U} - \omega^2 \boldsymbol{\rho}_{eff} \cdot \boldsymbol{U} \tag{2}$$

where $\boldsymbol{\Sigma}$ is the stress amplitude tensor; $\boldsymbol{C}$ is the elastic tensor; $\boldsymbol{U}$ is the displacement amplitude vector; $\boldsymbol{S}$ is a third-order tensor; and the effective density tensor $\boldsymbol{\rho}_{eff}$ is the function of wave angular frequency $\omega$. The complex-valued counterparts of Eqs. (1) and (2) in the time domain are given as [14]

$$\boldsymbol{\sigma} = \boldsymbol{C} : \nabla \boldsymbol{u} - \frac{1}{\mathrm{i}\omega} \boldsymbol{S} \cdot \dot{\boldsymbol{u}} \tag{3}$$

$$\nabla \cdot \boldsymbol{\sigma} = -\frac{1}{\mathrm{i}\omega} \boldsymbol{S}^{\mathrm{T}} : \nabla \dot{\boldsymbol{u}} + \boldsymbol{\rho}_{eff} \cdot \ddot{\boldsymbol{u}} \tag{4}$$

where $\boldsymbol{u}$ is the displacement vector; $\boldsymbol{\sigma}$ is the stress tensor; and the overhead dot denotes the derivative with time.

The key point of this paper is to propose a general approach to investigate the form-invariance of wave equations in the global Cartesian coordinate system. The starting point is the fact that correct wave equations must be form-invariant in Lagrange coordinate system, which could not be Cartesian, according to the definition of tensor. Then the wave equation in the physical space is transformed from Lagrange coordinate system into Euler coordinate system, which is global Cartesian, to discuss its form-invariance. Up to this step, the field variables in the physical space have nothing to do with their counterparts in the virtual space. However, if the wave is required to propagate along a special trajectory, one can obtain the corresponding material properties with specific relations between the field variables in both spaces. In this way, one can clearly find out that Maxwell equations and sound equations are form-invariant and locally accurate for inhomogeneous media, while the traditional elastodynamic equations are not. In addition, a local form of Willis equations in the time domain can be naturally obtained, which is form-invariant and real-valued with symmetric elastic tensor. Its accuracy in inhomogeneous media is demonstrated by numerical examples.

## 2. The investigation of form-invariance

The transformation method can be illustrated in Figure 1, in which an object in the virtual space $\Omega$ is transformed into the physical space $\Omega'$ via a mapping $\boldsymbol{x}' = \boldsymbol{x}'(\boldsymbol{x})$. This process can be described in Euler representation: a material point at location $\boldsymbol{X} = x^i \boldsymbol{e}_i$ (indices $i$ = 1, 2, 3 and follow the rule of Einstein summation) moves to location $\boldsymbol{x}' = x'^i \boldsymbol{e}'_i$ in the global Cartesian coordinate system with the constant basis $\boldsymbol{e}_i = \boldsymbol{e}'_i$ ($i$ = 1, 2, 3). This process can also be described in Lagrange representation: the material point $\boldsymbol{X} = X^I {}_0\boldsymbol{g}_I$ moves to location $\boldsymbol{X} = X^I {}_t\boldsymbol{g}_I$ in local material coordinate system, whose covariant basis changes from ${}_0\boldsymbol{g}_I$ to ${}_t\boldsymbol{g}_I$ ($I$ = 1, 2, 3) after the coordinate transformation. In the virtual space, the local material coordinate systems at all points are the same as the Cartesian coordinate system in the fixed global space, i.e., ${}_0\boldsymbol{g}_1 = \boldsymbol{e}_1$, ${}_0\boldsymbol{g}_2 = \boldsymbol{e}_2$ and ${}_0\boldsymbol{g}_3 = \boldsymbol{e}_3$. However, in the transformed physical space, the local material coordinate system varies from point to point and is not orthogonal, generally.

Since $\mathrm{d}_t\boldsymbol{r} = \mathrm{d}X^I {}_t\boldsymbol{g}_I = \mathrm{d}x'^i \boldsymbol{e}'_i = \mathrm{d}x'^i \boldsymbol{e}_i$, one obtains the relation:

$$ {}_t\boldsymbol{g}_I = \frac{\partial x'^i}{\partial X^I}\boldsymbol{e}_i = F_I^{i'}\boldsymbol{e}_i \tag{5}$$

where $F_I^{i'} = \dfrac{\partial x'^i}{\partial X^I}$ is the transformation tensor between the global Cartesian coordinate system and the local material coordinate system for the configuration in the physical space. It can be generalized to the two-point deformation tensor $\boldsymbol{F} = F_i^{i'}\boldsymbol{e}'_i\boldsymbol{e}^i = \dfrac{\partial x'^i}{\partial x^i}\boldsymbol{e}_i\boldsymbol{e}^i$ that connects the coordinate increments between the physical space and the virtual space, i.e., $\mathrm{d}\boldsymbol{x}' = \boldsymbol{F}\cdot\mathrm{d}\boldsymbol{x}$. In the following, superscripts denote contravariant components and subscripts denote covariant components.

One of the intrinsic properties of tensor equations is the form-invariance after arbitrary coordinate transformation. Therefore, in the following investigations, wave equations are firstly written in their general tensor forms, which should be the same in both the virtual space and the physical space. Then, these equations in the physical space $\Omega'$ are represented in the global Cartesian coordinate system with the help of the transformation tensor given in Eq. (5). This gives a general approach to check if they still have the same forms as those in the virtual space. Up to this step, one does not need any assumption of the relations between field variables in the virtual space and the physical space. However, specific relations do need to obtain effective material parameters for wave controlling.

### 2.1. *Electromagnetic wave*

Maxwell field equations can be written as:

$$\nabla \times \boldsymbol{E} + \dot{\boldsymbol{B}} = \mathbf{0} \tag{6a}$$

$$\nabla \times \boldsymbol{H} = \boldsymbol{J} + \dot{\boldsymbol{D}} \tag{6b}$$

$$\nabla \cdot \boldsymbol{D} = \rho_c \tag{6c}$$

$$\nabla \cdot \boldsymbol{B} = 0 \tag{6d}$$

where $\boldsymbol{E}$ is the electric field intensity vector; $\boldsymbol{B}$ is the magnetic induction vector; $\boldsymbol{H}$ is the magnetic field intensity vector; $\boldsymbol{D}$ is the electric displacement vector; $\boldsymbol{J}$ is the current density vector; $\rho_c$ is the volume charge density.

And the constitutive relations are:

$$\boldsymbol{D} = \boldsymbol{\varepsilon} \cdot \boldsymbol{E} \tag{7a}$$

$$\boldsymbol{B} = \boldsymbol{\mu} \cdot \boldsymbol{H} \tag{7b}$$

$$\boldsymbol{J} = \boldsymbol{s} \cdot \boldsymbol{E} \tag{7c}$$

where $\boldsymbol{\varepsilon}$ is the electric permittivity tensor; $\boldsymbol{\mu}$ is the magnetic permeability tensor; and $\boldsymbol{s}$ is the electric conductivity tensor.

In the virtual space $\Omega$, Eq. (7*a*) can be written in the global Cartesian coordinate system as:

$$ {}_0D^i = {}_0\varepsilon\delta^{ij}\, {}_0E_j \tag{8}$$

where $\delta^{ij}$ is the Kronecker delta.

In the physical space $\Omega'$, Eq. (7*a*) can be written in the local non-Cartesian coordinate system as:

$$ {}_tD^I = {}_t\varepsilon^{IJ}\, {}_tE_J \tag{9}$$

which can be transformed into the global Cartesian coordinate system by using Eq. (5):

$$\begin{gathered} F^I_{i'}\, {}_tD^{i'} = F^I_{i'} F^J_{j'}\, {}_t\varepsilon^{i'j'} F^{k'}_J\, {}_tE_{k'} \\ {}_tD^{i'} = {}_t\varepsilon^{i'j'}\, {}_tE_{j'} \end{gathered} \tag{10}$$

which has the same form as Eq. (8). Similarly, other constitutive equations are form-invariant in the global Cartesian coordinate system.

In the virtual space $\Omega$, Eq. (6*a*) can be written in the global Cartesian coordinate system as:

$$ {}_0e^{ijk}\, {}_0E_{j,i} + {}_0\dot{B}^k = 0 \tag{11}$$

where ${}_0e^{ijk}$ is the Levi-Civita symbol and the subscript $_{,i}$ denotes the partial derivative with respect to $x^i$.

In the physical space $\Omega'$, Eq. (6*a*) can be written in the local non-Cartesian coordinate system as:

$$\begin{gathered} {}_t\mathscr{E}^{IJK}\, {}_tE_{J;I} + {}_t\dot{B}^K = 0 \\ {}_t\mathscr{E}^{IJK}\left({}_tE_{J,I} - {}_tE_M\, {}_t\Gamma^M_{JI}\right) + {}_t\dot{B}^K = 0 \\ {}_t\mathscr{E}^{IJK}\, {}_tE_{J,I} + {}_t\dot{B}^K = 0 \end{gathered} \tag{12}$$

where ${}_t\mathscr{E}^{IJK}$ is the contravariant component of Eddington tensor; the subscript $_{;I}$ denotes the covariant derivative with respect to $X^I$; ${}_t\Gamma^M_{JI} = {}_t\Gamma^M_{IJ}$ is the Christoffel symbol and ${}_t\mathscr{E}^{IJK}\, {}_t\Gamma^M_{JI} = 0$ since ${}_t\mathscr{E}^{IJK} = -{}_t\mathscr{E}^{JIK}$. With $F^J_{j'}\, {}_t\mathscr{E}^{i'j'k'} F^{r'}_{J,i'} = {}_t\mathscr{E}^{i'j'k'} F^J_{j'} F^{r'}_{J,I} F^I_{i'} = {}_t\mathscr{E}^{i'j'k'} F^J_{i'} F^{r'}_{J,I} F^I_{j'} = 0$ and ${}_t\mathscr{E}^{i'j'k'} = {}_te^{i'j'k'}$ in the global Cartesian coordinate system being the case, Eq. (12) can be transformed into the global Cartesian coordinate system by using Eq. (5):

$$
\begin{gathered}
F_{i'}^I F_{j'}^J F_{k'}^K \, {}_t\mathscr{E}^{i'j'k'} F_I^{s'} \left(F_J^{r'} \, {}_tE_{r'}\right)_{,s'} + F_{k'}^K \, {}_t\dot{B}^{k'} = 0 \\
F_{j'}^J \, {}_t\mathscr{E}^{i'j'k'} \left(F_{J,i'}^{r'} \, {}_tE_{r'} + F_J^{r'} \, {}_tE_{r',i'}\right) + {}_t\dot{B}^{k'} = 0 \\
{}_te^{i'j'k'} \, {}_tE_{j',i'} + {}_t\dot{B}^{k'} = 0
\end{gathered} \tag{13}
$$

which has the same form as Eq. (11).

In the virtual space $\Omega$ , Eq. (6*c*) can be written in the global Cartesian coordinate system as:

$$
{}_0D_{,i}^i = {}_0\rho_c \tag{14}
$$

In the physical space $\Omega'$ , Eq. (6*c*) can be written in the local non-Cartesian coordinate system as:

$$
\begin{gathered}
{}_tD_{;I}^I = {}_t\rho_c \\
{}_tD_{,I}^I + {}_tD^J \, {}_t\Gamma_{JI}^I = {}_t\rho_c
\end{gathered} \tag{15}
$$

Noticing ${}_t\Gamma_{JI}^I = -F_{r',s'}^I F_J^{r'} F_I^{s'}$ and using Eq. (5), Eq. (15) can be transformed into the global Cartesian coordinate system as:

$$
\begin{gathered}
F_I^{j'} \left(F_{i'}^I \, {}_tD^{i'}\right)_{,j'} - F_{i'}^J \, {}_tD^{i'} F_{r',s'}^I F_J^{r'} F_I^{s'} = {}_t\rho_c \\
{}_tD_{,i'}^{i'} = {}_t\rho_c
\end{gathered} \tag{16}
$$

which has the same form as Eq. (14). Actually, since Eq. (6c) is a scalar equation, Eq. (16) is a natural result.

Similarly, other field equations are also form-invariant in the global Cartesian coordinate system.

In the above proof, there is not any assumption of the relations between field variables. However, certain assumptions do need to obtain effective material properties for a specific requirement of how to steer the wave front. For example, substituting Eq. (7*b*) into Eq. (11) obtains:

$$
{}_0e^{ijk} \, {}_0E_{j,i} + {}_0\mu\delta^{kl} \, {}_0\dot{H}_l = 0 \tag{17}
$$

And Eq. (13) can be written as:

$$
{}_te^{i'j'k'} \, {}_tE_{j',i'} + {}_t\mu^{k'l'} \, {}_t\dot{H}_{l'} = 0 \tag{18}
$$

If ${}_tE_{j'} = F_{j'}^j \, {}_0E_j$ , ${}_tH_{l'} = F_{l'}^l \, {}_0H_l$ is required, and ${}_te^{i'j'k'} F_{j',i'}^j = 0$ and ${}_te^{i'j'k'} F_{i'}^i F_{j'}^j F_{k'}^k = \frac{1}{{}_tJ} \, {}_0e^{ijk}$ are observed, where ${}_tJ = \det\left(F_i^{i'}\right)$ represents the volumetric change ratio after the coordinate transformation from the virtual space to the physical space, Eq. (18) becomes:

$$
\begin{gathered}
{}_te^{i'j'k'} \left(F_{j'}^j \, {}_0E_j\right)_{,i'} + {}_t\mu^{k'l'} F_{l'}^l \, {}_0\dot{H}_l = 0 \\
{}_te^{i'j'k'} F_{i'}^i F_{j'}^j \, {}_0E_{j,i} + {}_t\mu^{k'l'} F_{l'}^l \, {}_0\dot{H}_l = 0 \\
\frac{1}{{}_tJ} \, {}_0e^{ijk} \, {}_0E_{j,i} + {}_t\mu^{k'l'} F_{k'}^k F_{l'}^l \, {}_0\dot{H}_l = 0
\end{gathered} \tag{19}
$$

Comparing Eqs. (17) and (19), obtains ${}_t\mu^{k'l'} = \frac{{}_0\mu \, {}_tF_k^{k'} F_l^{l'} \delta^{kl}}{{}_tJ} = \frac{{}_0\mu \, {}_tg^{k'l'}}{{}_tJ}$ , where ${}_tg^{k'l'} = F_k^{k'} F_k^{l'}$ is the metric tensor. This is a well known result [9].

### 2.2. *Sound wave*

For sound wave, the constitutive and field equations are:

$$
p + K\nabla \cdot \boldsymbol{u} = 0 \tag{20}
$$

$$
\nabla p + \boldsymbol{\rho} \cdot \ddot{\boldsymbol{u}} = \boldsymbol{0} \tag{21}
$$

where $p$ is the pressure; $K$ is the bulk stiffness; and $\boldsymbol{\rho}$ is the mass density tensor.

In the virtual space $\Omega$, Eqs. (20) and (21) can be written in the global Cartesian coordinate system as:

$$ {}_0p + {}_0K\, {}_0u^i_{,i} = 0 \tag{22}$$

$$ {}_0p_{,j} + {}_0\rho\delta_{ji}\, {}_0\ddot{u}^i = 0 \tag{23}$$

In the physical space $\Omega'$, since Eq. (20) is a scalar equation, it can be directly written (or follows the deductions of Eq. (15) through Eq. (16)) in the global Cartesian coordinate system as:

$$ {}_tp + {}_tK\, {}_tu^{i'}_{,i'} = 0 \tag{24}$$

which has the same form as Eq. (22).

In the physical space $\Omega'$, Eq. (21) can be written in the local non-Cartesian coordinate system as:

$$ {}_tp_{,J} + {}_t\rho_{JI}\, {}_t\ddot{u}^I = 0 \tag{25}$$

which can be transformed into the global Cartesian coordinate system by using Eq. (5):

$$\begin{gathered} F^{j'}_J\, {}_tp_{,j'} + F^{r'}_I F^{s'}_J\, {}_t\rho_{s'r'} F^I_{i'}\, {}_t\ddot{u}^{i'} = 0 \\ {}_tp_{,j'} + {}_t\rho_{j'i'}\, {}_t\ddot{u}^{i'} = 0 \end{gathered} \tag{26}$$

which has the same form as Eq. (23).

In the above proof, there is not any assumption of the relation between field variables in the virtual and the physical spaces. If require ${}_tp = {}_0p$ and ${}_tu^{i'} = \frac{1}{{}_tJ} F^{i'}_i\, {}_0u^i$, and use Piola identity $\left(\frac{1}{{}_tJ} F^{i'}_i\right)_{,i'} = 0$, Eqs. (24) and (26) become:

$$\begin{aligned} {}_0p + {}_tK\left(\frac{1}{{}_tJ} F^{i'}_i\, {}_0u^i\right)_{,i'} &= {}_0p + \frac{{}_tK}{{}_tJ}\, {}_0u^i_{,i} \\ &= 0 \end{aligned} \tag{27}$$

$$\begin{gathered} F^j_{j'}\, {}_0p_{,j} + {}_t\rho_{j'i'} \frac{1}{{}_tJ} F^{i'}_i\, {}_0\ddot{u}^i = 0 \\ {}_0p_{,j} + {}_t\rho_{i'j'} \frac{F^{j'}_j F^{i'}_i}{{}_tJ}\, {}_0\ddot{u}^i = 0 \end{gathered} \tag{28}$$

Comparing Eq. (22) with Eq. (27) and Eq. (23) with Eq. (28), obtains ${}_tK = {}_tJ\, {}_0K$ and ${}_t\rho_{i'j'} = {}_tJ\, {}_0\rho\delta_{ij} F^i_{i'} F^j_{j'} = {}_tJ\, {}_0\rho\, {}_tg_{i'j'}$, which coincides with the results in [22].

### 2.3. *Elastic wave*

The conventional linear elastodynamic wave equations are:

$$ \boldsymbol{\sigma} = \boldsymbol{C} : \nabla \boldsymbol{u} \tag{29}$$

$$ \nabla\cdot\boldsymbol{\sigma} + \boldsymbol{f} = \boldsymbol{\rho}\cdot\ddot{\boldsymbol{u}} \tag{30}$$

where $\boldsymbol{f}$ is the body force vector. Unfortunately, these equations are not form-invariant under arbitrary coordinate transformation [9]. And this paper tries to give new elastodynamic equations that are real-valued and form-invariant.

Since only linear elastic wave is considered in this paper, the relations between field variables in the virtual space and in the physical space can be generally expressed as:

$$ {}_0\sigma^{ij} = A^i_{i'} A^j_{j'}\, {}_t\sigma^{i'j'}, \quad {}_0u_k = B^{k'}_k\, {}_tu_{k'} \quad \text{and} \quad {}_0f^i = P^i_{k'}\, {}_tf^{k'} \tag{31}$$

where $A_{i'}^{i}$, $B_{k}^{k'}$ and $P_{k'}^{i}$ are arbitrary gauges [14].

Substituting Eq. (31) into Eq. (29), gives:

$$A_{i'}^{i}A_{j'}^{j}\,{}_{t}\sigma^{i'j'} = {}_{0}C^{ijkl}F_{l}^{l'}\left(B_{k}^{k'}\,{}_{t}u_{k'}\right)_{,l'}$$

$${}_{t}\sigma^{i'j'} = {}_{0}C^{ijkl}\left(A^{-1}\right)_{i}^{i'}\left(A^{-1}\right)_{j}^{j'}F_{l}^{l'}B_{k}^{k'}\,{}_{t}u_{k',l'} + {}_{0}C^{ijkl}\left(A^{-1}\right)_{i}^{i'}\left(A^{-1}\right)_{j}^{j'}F_{l}^{l'}B_{k,l'}^{k'}\,{}_{t}u_{k'}$$

I.e., the constitutive equation could be:

$$\boldsymbol{\sigma} = \boldsymbol{C} : \nabla\boldsymbol{u} + \boldsymbol{S}\cdot\boldsymbol{u} \tag{32}$$

where $\boldsymbol{S}$ is a third-order tensor.

Substituting Eq. (31) into Eq. (30), yields:

$$F_{j}^{r'}\left(A_{i'}^{i}A_{j'}^{j}\,{}_{t}\sigma^{i'j'}\right)_{,r'} + P_{k'}^{i}\,{}_{t}f^{k'} = {}_{0}\rho\delta^{ik}B_{k}^{r'}\,{}_{t}\ddot{u}_{r'}$$

$${}_{t}\sigma_{,j'}^{i'j'} + F_{j'}^{j}\left(A^{-1}\right)_{i}^{i'}\left(A^{-1}\right)_{j}^{j'}P_{k'}^{i}\,{}_{t}f^{k'} = F_{j'}^{j}\left(A^{-1}\right)_{i}^{i'}\left(A^{-1}\right)_{j}^{j'}\,{}_{0}\rho\delta^{ik}B_{k}^{r'}\,{}_{t}\ddot{u}_{r'} - \left(A^{-1}\right)_{i}^{r'}\left(A^{-1}\right)_{j}^{s'}\left(A_{r'}^{i}A_{s'}^{j}\right)_{,j'}\,{}_{t}\sigma^{i'j'}$$

By referring to Eq. (32), the equilibrium equation could be obtained from the above equation:

$$\nabla\cdot\boldsymbol{\sigma} + \boldsymbol{f} = \boldsymbol{\rho}\cdot\ddot{\boldsymbol{u}} + \boldsymbol{D} : \nabla\boldsymbol{u} + \boldsymbol{K}\cdot\boldsymbol{u} \tag{33}$$

where $\boldsymbol{D}$ is a third-order tensor and $\boldsymbol{K}$ is a two-order tensor.

Compared with Eqs. (29) and (30), Eqs. (32) and (33) have three additional terms. The form-invariant properties of Eqs. (32) and (33) will be proven in the following.

In the virtual space $\Omega$, Eqs. (32) and (33) can be written in the global Cartesian coordinate system as:

$${}_{0}\sigma^{ij} = {}_{0}C^{ijkl}\,{}_{0}u_{k,l} + {}_{0}S^{ijk}\,{}_{0}u_{k} \tag{34}$$

$${}_{0}\sigma_{,j}^{ij} + {}_{0}f^{i} = {}_{0}\rho\delta^{ij}\,{}_{0}\ddot{u}_{j} + {}_{0}D^{ikl}\,{}_{0}u_{k,l} + {}_{0}K^{ij}\,{}_{0}u_{j} \tag{35}$$

If the material is homogeneous, ${}_{0}S^{ijk} = 0$, ${}_{0}D^{ikl} = 0$ and ${}_{0}K^{ij} = 0$. However, these constraints are not generally necessary in the following proof of form-invariance.

In the physical space $\Omega'$, Eq. (32) can be written in the local non-Cartesian coordinate system as:

$${}_{t}\sigma^{IJ} = {}_{t}C^{IJKL}\,{}_{t}u_{K;L} + {}_{t}S^{IJK}\,{}_{t}u_{K} \tag{36}$$

which can be transformed into the global Cartesian coordinate system by using Eq. (5):

$$\begin{aligned} F_{i'}^{I}F_{j'}^{J}\,{}_{t}\sigma^{i'j'} &= F_{i'}^{I}F_{j'}^{J}F_{k'}^{K}F_{l'}^{L}\,{}_{t}C^{i'j'k'l'}F_{K}^{r'}F_{L}^{s'}\,{}_{t}u_{r',s'} + F_{i'}^{I}F_{j'}^{J}F_{k'}^{K}\,{}_{t}S^{i'j'k'}F_{K}^{r'}\,{}_{t}u_{r'} \\ {}_{t}\sigma^{i'j'} &= {}_{t}C^{i'j'k'l'}\,{}_{t}u_{k',l'} + {}_{t}S^{i'j'k'}\,{}_{t}u_{k'} \end{aligned} \tag{37}$$

which has the same form as Eq. (34).

In the physical space $\Omega'$, Eq. (33) can be written in the local non-Cartesian coordinate system as:

$${}_{t}\sigma_{;J}^{IJ} + {}_{t}f^{I} = {}_{t}\rho^{IJ}\,{}_{t}\ddot{u}_{J} + {}_{t}D^{IKL}\,{}_{t}u_{K;L} + {}_{t}K^{IJ}\,{}_{t}u_{J} \tag{38}$$

which can be transformed into global Cartesian coordinate system by using Eq. (5):

$$\begin{aligned} F_{i'}^{I}\,{}_{t}\sigma_{,j'}^{i'j'} + F_{i'}^{I}\,{}_{t}f^{i'} &= F_{i'}^{I}F_{j'}^{J}\,{}_{t}\rho^{i'j'}F_{J}^{k'}\,{}_{t}\ddot{u}_{k'} + F_{i'}^{I}F_{k'}^{K}F_{l'}^{L}\,{}_{t}D^{i'k'l'}F_{K}^{r'}F_{L}^{s'}\,{}_{t}u_{r',s'} + F_{i'}^{I}F_{j'}^{J}\,{}_{t}K^{i'j'}F_{J}^{r'}\,{}_{t}u_{r'} \\ {}_{t}\sigma_{,j'}^{i'j'} + {}_{t}f^{i'} &= {}_{t}\rho^{i'j'}\,{}_{t}\ddot{u}_{j'} + {}_{t}D^{i'k'l'}\,{}_{t}u_{k',l'} + {}_{t}K^{i'j'}\,{}_{t}u_{j'} \end{aligned} \tag{39}$$

which has the same form as Eq. (35).

In the above proof, there is no assumption of the relation between field variables in the virtual and the physical spaces. If ${}_{t}\sigma^{i'j'} = \frac{1}{{}_{t}J}F_{i}^{i'}F_{j}^{j'}\,{}_{0}\sigma^{ij}$ and ${}_{t}u_{i'} = F_{i'}^{i}\,{}_{0}u_{i}$ are required, Eq. (37) becomes:

$$
\begin{aligned}
\frac{1}{{}_tJ} F_i^{i'} F_j^{j'} {}_0\sigma^{ij} &= {}_tC^{i'j'k'l'} F_{l'}^{l} \left( F_{k'}^{r} {}_0u_r \right)_{,l} + {}_tS^{i'j'k'} F_{k'}^{r} {}_0u_r \\
{}_0\sigma^{ij} &= {}_tJ F_{i'}^{i} F_{j'}^{j} F_{l'}^{l} {}_tC^{i'j'k'l'} \left( F_{k',l}^{r} {}_0u_r + F_{k'}^{r} {}_0u_{r,l} \right) + {}_tJ F_{i'}^{i} F_{j'}^{j} F_{k'}^{r} {}_tS^{i'j'k'} {}_0u_r \\
{}_0\sigma^{ij} &= {}_tJ F_{i'}^{i} F_{j'}^{j} F_{k'}^{k} F_{l'}^{l} {}_tC^{i'j'k'l'} {}_0u_{k,l} + {}_tJ F_{i'}^{i} F_{j'}^{j} \left( {}_tC^{i'j'k'l'} F_{l'}^{l} F_{k',l}^{r} + F_{k'}^{r} {}_tS^{i'j'k'} \right) {}_0u_r
\end{aligned}
\tag{40}
$$

With ${}_0\sigma^{ij} = {}_0C^{ijkl} {}_0u_{k,l}$ in homogeneous media being the case, Eq. (39) becomes:

$$
\begin{aligned}
\left( \frac{1}{{}_tJ} F_i^{i'} F_j^{j'} {}_0\sigma^{ij} \right)_{,j'} + {}_tf^{i'} &= {}_t\rho^{i'j'} F_{j'}^{j} {}_0\ddot{u}_j + {}_tD^{i'k'l'} F_{l'}^{l} \left( F_{k'}^{k} {}_0u_k \right)_{,l} + {}_tK^{i'j'} F_{j'}^{k} {}_0u_k \\
\frac{1}{{}_tJ} F_j^{j'} \left( F_i^{i'} {}_0\sigma^{ij} \right)_{,j'} + {}_tf^{i'} &= {}_t\rho^{i'j'} F_{j'}^{j} {}_0\ddot{u}_j + {}_tD^{i'k'l'} F_{l'}^{l} \left( F_{k',l}^{k} {}_0u_k + F_{k'}^{k} {}_0u_{k,l} \right) + {}_tK^{i'j'} F_{j'}^{k} {}_0u_k \\
{}_0\sigma^{ij}_{,j} + {}_tJ F_{i'}^{i} {}_tf^{i'} &= {}_tJ\, {}_t\rho^{i'j'} F_{i'}^{i} F_{j'}^{j} {}_0\ddot{u}_j + {}_tJ\, {}_tD^{i'k'l'} F_{i'}^{i} F_{l'}^{l} \left( F_{k',l}^{k} {}_0u_k + F_{k'}^{k} {}_0u_{k,l} \right) + {}_tJ\, {}_tK^{i'j'} F_{i'}^{i} F_{j'}^{k} {}_0u_k \\
&\quad - F_{i'}^{i} F_j^{j'} F_{i,j'}^{i'} {}_0C^{ijkl} {}_0u_{k,l} \\
&= {}_tJ\, {}_t\rho^{i'j'} F_{i'}^{i} F_{j'}^{j} {}_0\ddot{u}_j + F_{i'}^{i} \left( {}_tJ\, {}_tD^{i'k'l'} F_{l'}^{l} F_{k'}^{k} - F_j^{j'} F_{i,j'}^{i'} {}_0C^{ijkl} \right) {}_0u_{k,l} \\
&\quad + {}_tJ\, {}_tF_{i'}^{i} \left( {}_tD^{i'k'l'} F_{l'}^{l} F_{k',l}^{k} + {}_tK^{i'j'} F_{j'}^{k} \right) {}_0u_k
\end{aligned}
\tag{41}
$$

Comparing Eq. (34) with Eq. (40), and noticing ${}_0S^{ijk} = 0$ in homogeneous media, yields

$$
{}_tC^{i'j'k'l'} = \frac{1}{{}_tJ} F_i^{i'} F_j^{j'} F_k^{k'} F_l^{l'} {}_0C^{ijkl} \tag{42a}
$$

$$
{}_tS^{i'j'k'} = \frac{1}{{}_tJ} F_i^{i'} F_j^{j'} F_{k,l}^{k'} {}_0C^{ijkl} \tag{42b}
$$

Comparing Eq. (35) with Eq. (41), and noticing ${}_0D^{ikl} = 0$ and ${}_0K^{ij} = 0$ in homogeneous media, gives

$$
{}_tf^{i'} = \frac{1}{{}_tJ} F_i^{i'} {}_0f^{i} \tag{42c}
$$

$$
{}_t\rho^{i'j'} = \frac{1}{{}_tJ} {}_0\rho\, {}_tg^{i'j'} \tag{42d}
$$

$$
{}_tD^{i'k'l'} = \frac{1}{{}_tJ} F_l^{l'} F_k^{k'} F_{i,j}^{i'} {}_0C^{ijkl} = {}_tS^{k'l'i'} \tag{42e}
$$

$$
{}_tK^{i'j'} = \frac{1}{{}_tJ} F_{i,j}^{i'} F_{k,l}^{j'} {}_0C^{ijkl} \tag{42f}
$$

The effective material parameters in Eq. (42) coincide with the results in [9] and the corresponding stress tensor maintains its symmetry. In this case, Eqs. (32) and (33) could be regarded as the real-valued counterparts of Eqs. (1) and (2) [9] in the time domain.

### 2.4. *Discussions*

So far this paper uses a unified approach to investigate the form-invariance of wave equations. The merit of this approach is not assuming any relation between the field variables in the virtual and the physical spaces. Thus, it can reveal the intrinsic properties of wave equations.

The results show that Maxwell equations and sound equations are form-invariant, so that they can accurately describe the wave propagation in inhomogeneous media. However, traditional elastodynamic equations do not have this property [9] and are accurate only for homogeneous media. This is probably because electro-magnetic waves are pure transverse waves; sound waves are pure longitudinal waves; while elastic waves are much more complex. The coupling of the transverse wave and the longitudinal wave results in additional terms in elastodynamic wave equations for inhomogeneous media. If substituting Eq. (32) into Eq. (33), one can

obtain:

$$\boldsymbol{\rho}\cdot\ddot{\boldsymbol{u}}+\nabla\cdot\left(-\boldsymbol{C}:\nabla\boldsymbol{u}\right)-\boldsymbol{f}+\nabla\cdot\left(-\boldsymbol{S}\cdot\boldsymbol{u}\right)+\boldsymbol{D}:\nabla\boldsymbol{u}+\boldsymbol{K}\cdot\boldsymbol{u}=\boldsymbol{0} \tag{43}$$

The first three terms in the above equation are the same as those in traditional Lamé-Navier equation. Other three additional terms exist only for inhomogeneous media and have clear physical meanings: $\nabla\cdot\left(-\boldsymbol{S}\cdot\boldsymbol{u}\right)$ represents the conservative flux convection; $\boldsymbol{D}:\nabla\boldsymbol{u}$ represents the convection; and $\boldsymbol{K}'\cdot\boldsymbol{u}'$ represents the absorption. In addition, the third-order tensor $\boldsymbol{S}$ in Eq. (32) could be regarded as the gradient of pre-stresses, which are the cause of inhomogeneity.

Although the coordinate transformation method is used in the above investigation, the obtained results could be more general. Remember that Willis had given the similar results over thirty years ago using another approach [16, 17]. Eqs. (32) and (33) with material properties in Eq. (42) could be regarded as local versions of Willis equations if the microstructure of the material is sufficiently small compared with the interested wave length [9]. And generally, the stress tensor in Eqs. (32) and (33) is not necessarily symmetric as pointed out by Norris and Shuvalov [14].

It seems that effective material parameters could be obtained for any specific relations between field variables in the virtual and the physical spaces. However, a physical implementation should not violate the conservation of energy and mass [23]. For example, to obtain the effective parameters in Eq. (42), one should not remove ${}_tJ$ from ${}_t\sigma^{i'j'}=\dfrac{1}{{}_tJ}F_i^{i'}F_j^{j'}\,{}_0\sigma^{ij}$. In a word, one can only steer the wave front along some special but not arbitrary trajectories.

## 3. Numerical examples

To demonstrate the validity of the new elastodynamic Eqs. (32) and (33), we have designed a perfect plane wave rotator and an approximate plane wave cloak as follows.

### 3.1. *A perfect wave rotator*

As Figure 2 shows, this wave rotator is a ring structure with an inner boundary of radius $a$ = 0.05m and out outer boundary of radius $b$ = 0.15m. It is embedded in an isotropic and homogeneous background with the Young's modulus $E=2\times10^{11}$ Pa, the Poisson's ratio $\nu=0.3$ and the density $\rho=7800$ Kg/m$^3$. When a plane wave incidents from left of this rotator, it will rotate $\alpha$ angle and then recovers its original form when it leaves this region. This device can be designed by the following mappings:

$$r'=r \quad \text{and} \quad \theta'=\theta+f\left(r\right)\alpha \tag{44}$$

where $f\left(r\right)=Ar^3+Br^2+Cr+D$ that satisfies $f\left(a\right)=1$ and $f\left(b\right)=0$. In this case, the array of deformation gradient in the polar coordinate system is

$$\left[F_i^{i'}\right]=\begin{bmatrix}\dfrac{\partial r'}{\partial r} & \dfrac{\partial r'}{r\partial\theta}\\ r'\dfrac{\partial\theta'}{\partial r} & \dfrac{r'}{r}\dfrac{\partial\theta'}{\partial\theta}\end{bmatrix}=\begin{bmatrix}1 & 0\\ \alpha r'\dfrac{\mathrm{d}f\left(r\right)}{\mathrm{d}r} & 1\end{bmatrix} \tag{45}$$

To meet the impedance-matching condition on both the inner and outer boundaries, requires $\dfrac{\mathrm{d}f\left(a\right)}{\mathrm{d}r}=\dfrac{\mathrm{d}f\left(b\right)}{\mathrm{d}r}=0$.

With the effective material parameters obtained from Eq. (42) (see Appendix A), numerical simulations can be implemented by solving Eqs. (32) and (33) (ignoring the body force) with the COMSOL Multiphysics™ software. As Figure 2 shows, incident plane elastic waves do rotate about the desired angle whether the rotator is designed by the new equations or the traditional equations (just remove ***S***, ***D*** and ***K*** from Eqs. (32) and (33)). However, only the rotators based on the new equations can smoothly recover the wave front without any abnormal scattering. This observation gives a clear evidence of the crucial help of the additional terms ***S***, ***D*** and ***K*** to the accurate description of elastic wave propagation in inhomogeneous media.

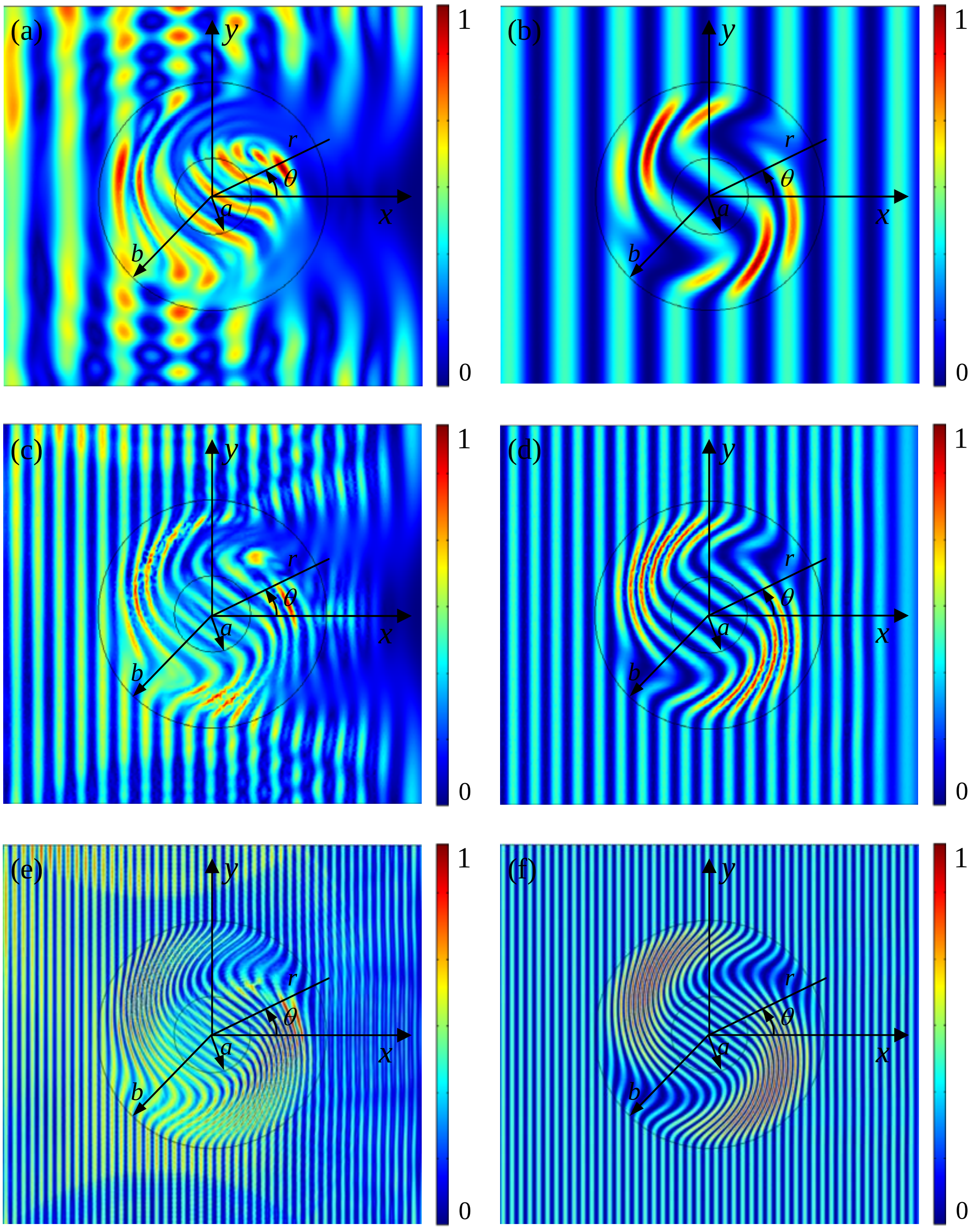

**Figure 2** (Colour on line) Snapshots of the normalized total displacement ( $\sqrt{u_x^2 + u_y^2}$ ) of plane elastic wave pass through wave rotators ( $\alpha = \frac{3\pi}{10}$ ): (a) Based on the traditional equations at 80kHz; (b) Based on the new equations at 80kHz; (c) Based on the traditional equations at 200kHz; (d) Based on the new equations at 200kHz; (e) Based on the traditional equations at 500kHz; (f) Based on the new equations at 500kHz.

In addition, it can be observed from Figures 2(a), 2(c) and 2(e) that the scattering is weaker and weaker with the increment of the wave frequency. This coincides with the observation in [23], where the elastic ray theory is used to explain that Lamé-Navier equation can be approximated by the eikonal equation for high frequency waves. Actually, at high frequencies the transverse wave and the longitudinal wave can be approximately decoupled [24]. Therefore, the eikonal equation is form-invariant. This can also be understood intuitively: with short wave length, it is difficult to *feel* the inhomogeneity.

### 3.2. *An approximate wave cloak*

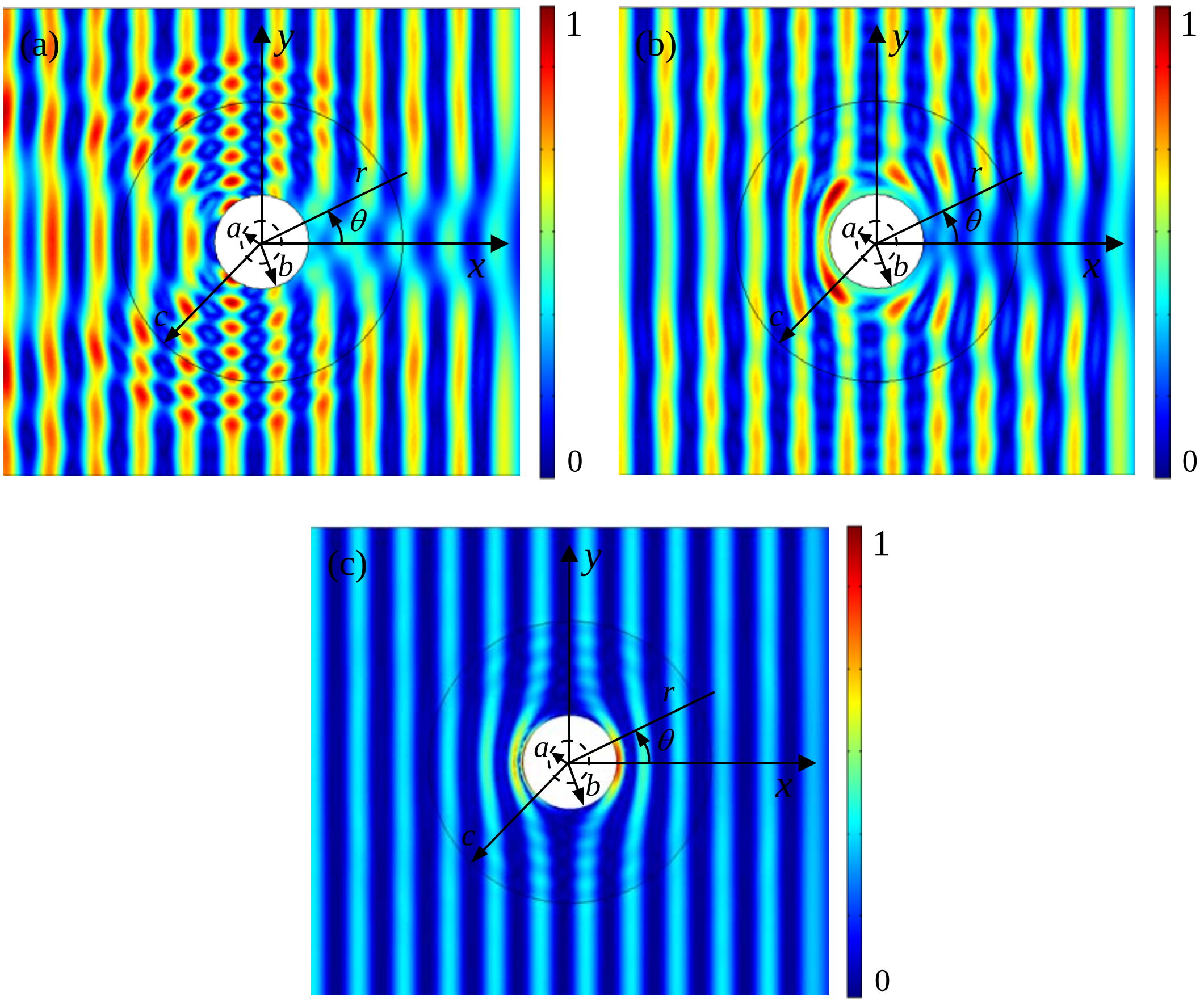

**Figure 3** (Colour on line) Snapshots of the normalized total displacement ($\sqrt{u_x^2+u_y^2}$) of plane elastic waves at 120kHz pass through voids: (a) Without cloak; (b) With the cloak based on the traditional equations; (c) With the cloak based on the new equations.

For further demonstrating the validity of the new elastodynamic equations, a plane approximate cloak is designed by mapping a circle with very small radius $a$ = 0.001m in the virtual domain into the inner boundary of radius $b$ = 0.05m, while keeping the outer boundary of radius $c$ = 0.15m unchanged (see Figure 3). The mapping functions are:

$$r' = f(r) \quad \text{and} \quad \theta' = \theta \tag{46}$$

where $f(r) = Ar^n + B$ that satisfies $f(a) = b$, $f(c) = c$. In this case, the array of deformation gradient in the polar coordinate system is

$$\left[F_i^{i'}\right] = \begin{bmatrix} \dfrac{\partial r'}{\partial r} & \dfrac{\partial r'}{r\partial\theta} \\ \dfrac{r'\partial\theta'}{\partial r} & \dfrac{r'\partial\theta'}{r\partial\theta} \end{bmatrix} = \begin{bmatrix} \dfrac{\mathrm{d}f(r)}{\mathrm{d}r} & 0 \\ 0 & \dfrac{r'}{r} \end{bmatrix} \tag{47}$$

So it requires $\frac{\mathrm{d}f(c)}{\mathrm{d}r}=1$ meeting the impedance-matching condition on the outer boundary. However, the impedance-matching condition cannot be generally satisfied on the inner boundary, because $\frac{\mathrm{d}f(a)}{\mathrm{d}r}\neq 1$ and $\left.\frac{r'}{r}\right|_{r=a}=\frac{b}{a}\neq 1$. Therefore, one can only expect approximate cloaking effect. In the following examples, the background material is the same as that of rotators shown in Figure 2. The detailed formulations of material parameters are given in Appendix B.

Figure 3(a) shows the strong scattering, when a plane wave passes through a void without any cloak. If the cloak designed by traditional equations is used, the incident wave front can bend around the central void but fail to recover completely behind the cloak. Therefore, a clear shadow can be identified behind the void (see Figure 3(b)). If the cloak designed by the new equations is used, the incident wave front can be recovered almost perfectly behind the cloak and only slight disturbances are observed around the inner boundary, where the impedance is different from that of the central void (see Figure 3(c)).

## 4. Concluding remarks

On the one hand, if a wave equation that can correctly describe physical phenomena in homogeneous and inhomogeneous media, it must necessarily have the same form in the global Cartesian coordinate system in both scenarios. On the other hand, if a wave equation can be written in tensor form, it must be form-invariant in local Lagrange coordinate system for any medium. Based on these two points, this paper investigates the form-invariance of wave equations by transforming them from the local Lagrange coordinate system to the global Cartesian coordinate system. This is a general approach that has nothing to do with the response of field variables. Thus, it can intrinsically reveal if a wave equation is accurate for inhomogeneous media.

The investigation shows that Maxwell equations and sound equations are accurate for inhomogeneous media but the traditional elastodynamic equations are not. These results coincide with the findings in literatures. Based on the obtained new elastodynamic equations, one can obtain local versions of Willis equations in time domain. The new equations have very clear physical meanings and their validity is illustrated numerically by some inverse scattering problems.

It is well known that an inverse problem should be discussed after the corresponding direct problem is well established and well-posed. Otherwise, large model errors will be presented. This is well demonstrated by the numerical examples presented in this paper. Further more, these findings also give a warning about relative applications, such as seismology and the nondestructive evaluation of composites, etc., where low frequency elastic waves do propagate in inhomogeneous media and the traditional elastodynamic equations are widely used.

## Acknowledgments

The author is grateful for the support from the National Science Foundation of China through grant 11272168. Thanks also go to Mr. Yi Xu for the discussion on numerical simulations.

## Appendices

### A. The design of perfect rotators

As Figure 2 shows, the following mappings are adopted to design the wave rotator:

$$r' = r \quad \text{and} \quad \theta' = \theta + f(r)\alpha \tag{A1}$$

where $f(r) = Ar^3 + Br^2 + Cr + D$ that satisfies $f(a) = 1$ and $f(b) = 0$. In addition, the impedance-matching condition requires $\frac{\mathrm{d}f(a)}{\mathrm{d}r} = \frac{\mathrm{d}f(b)}{\mathrm{d}r} = 0$, so that

$$A = -\frac{2}{(a-b)^3},\quad B = \frac{3(a+b)}{(a-b)^3},\quad C = -\frac{6ab}{(a-b)^3},\quad D = \frac{b^2(3a-b)}{(a-b)^3} \tag{A2}$$

Noticing $r = \sqrt{x^2 + y^2}$, $\cos\theta = \frac{x}{r}$ and $\sin\theta = \frac{y}{r}$ in the global Cartesian coordinate system, one obtains

$$x' = r'\cos\theta' = x\cos\left[f(r)\alpha\right] - y\sin\left[f(r)\alpha\right] \quad \text{and} \quad y' = r'\sin\theta' = y\cos\left[f(r)\alpha\right] + x\sin\left[f(r)\alpha\right] \tag{A3}$$

Then, the matrix of the deformation gradient can be easily obtained:

$$\left[F_i^{i'}\right] = \begin{bmatrix} cs(1-Hxy) - snHx^2 & -csHy^2 - sn(1+Hxy) \\ csHx^2 + sn(1-Hxy) & cs(1+Hxy) - snHy^2 \end{bmatrix} = \begin{bmatrix} P & Q \\ R & S \end{bmatrix} \tag{A4}$$

where $cs = \cos(f\alpha)$, $sn = \sin(f\alpha)$ and $H = \alpha(3Ar + 2B + C/r)$. And ${}_tJ = \det\left(F_i^{i'}\right) = 1$.

The matrix of metric tensor is

$$\left[{}_tg^{i'j'}\right] = \left[F_i^{i'}F_i^{j'}\right] = \begin{bmatrix} P^2 + Q^2 & PR + QS \\ PR + QS & R^2 + S^2 \end{bmatrix} \tag{A5}$$

with which one can easily obtain ${}_t\rho^{i'j'} = \frac{1}{{}_tJ}\,{}_0\rho\,{}_tg^{i'j'}$ from (42*d*).

Since ${}_0C^{ijkl} = \lambda\delta^{ij}\delta^{kl} + \mu\left(\delta^{ik}\delta^{jl} + \delta^{il}\delta^{jk}\right)$, according to (42*a*), the elastic tensor is:

$$\begin{aligned} {}_0C^{i'j'k'l'} &= \frac{1}{{}_tJ}\left[\lambda F_j^{i'}F_j^{j'}F_l^{k'}F_l^{l'} + \mu\left(F_k^{i'}F_k^{k'}F_l^{j'}F_l^{l'} + F_l^{i'}F_l^{l'}F_k^{j'}F_k^{k'}\right)\right] \\ &= \frac{1}{{}_tJ}\left[\lambda\,{}_tg^{i'j'}\,{}_tg^{k'l'} + \mu\left({}_tg^{i'k'}\,{}_tg^{j'l'} + {}_tg^{i'l'}\,{}_tg^{j'k'}\right)\right] \end{aligned} \tag{A6}$$

which can be written in matrix form as:

$$\begin{aligned} \left[{}_0C^{i'j'k'l'}\right] &= \begin{bmatrix} C^{1111} & C^{1122} & C^{1112} \\ & C^{2222} & C^{2212} \\ Sym. & & C^{1212} \end{bmatrix} \\ &= \begin{bmatrix} (\lambda+2\mu)\left(P^2+Q^2\right)^2 & \lambda\left(P^2+Q^2\right)\left(R^2+S^2\right) + 2\mu(PR+QS)^2 & (\lambda+2\mu)\left(P^2+Q^2\right)(PR+QS) \\ & (\lambda+2\mu)\left(R^2+S^2\right)^2 & (\lambda+2\mu)\left(R^2+S^2\right)(PR+QS) \\ Sym. & & (\lambda+\mu)(PR+QS)^2 + \mu\left(P^2+Q^2\right)\left(R^2+S^2\right) \end{bmatrix} \end{aligned} \tag{A7}$$

Similarly, the coupling terms can be calculated from (42*e*) as:

$$ {}_tD^{i'k'l'} = {}_tS^{k'l'i'} = \frac{1}{{}_tJ} F_k^{k'} F_l^{l'} F_{i,j}^{i'} C^{ijkl} = \frac{1}{{}_tJ} F_k^{k'} F_l^{l'} F_{i,j}^{i'} \left[ \lambda \delta^{ij} \delta^{kl} + \mu \left( \delta^{ik} \delta^{jl} + \delta^{il} \delta^{jk} \right) \right] = \frac{1}{{}_tJ} \left[ \lambda F_{i,i}^{i'} g^{k'l'} + \mu F_{i,j}^{i'} \left( F_i^{k'} F_j^{l'} + F_i^{l'} F_j^{k'} \right) \right] \tag{A8} $$

Since

$$ \frac{\partial \left[ F_i^{i'} \right]}{\partial x} = \begin{bmatrix} \Delta cs + \Phi sn & \Gamma cs + \Lambda sn \\ -\Phi cs + \Delta sn & -\Lambda cs + \Gamma sn \end{bmatrix} \text{ and } \frac{\partial \left[ F_i^{i'} \right]}{\partial y} = \begin{bmatrix} \Gamma cs + \Lambda sn & \Pi cs + \Theta sn \\ -\Lambda cs + \Gamma sn & -\Theta cs + \Pi sn \end{bmatrix} \tag{A9} $$

where $\Phi = H^2 x^2 y - x\left(3H + Kx^2\right)$ , $\Gamma = -H^2 x^2 y - x\left(H + Ky^2\right)$ , $\Lambda = H^2 xy^2 - y\left(H + Kx^2\right)$ , $\Pi = -H^2 xy^2 - y\left(3H + Ky^2\right)$, $\Theta = H^2 y^3 - x\left(H + Ky^2\right)$, $K = \left(3A/r - C/r^3\right)\alpha$, $\Delta = -H^2 x^3 - y\left(H + Kx^2\right)$, one can obtain

$$ \left[ D^{1k'l'} \right] = \begin{bmatrix} \Sigma cs + \Omega sn & \Xi cs + \Psi sn \\ \Xi cs + \Psi sn & \mathrm{X} cs + \mathrm{H} sn \end{bmatrix} \text{ and } \left[ D^{2k'l'} \right] = \begin{bmatrix} -\Omega cs + \Sigma sn & -\Psi cs + \Xi sn \\ -\Psi cs + \Xi sn & -\mathrm{H} cs + \mathrm{X} sn \end{bmatrix} \tag{A10} $$

where $\Sigma = \lambda\left(\Pi + \Delta\right)\left(P^2 + Q^2\right) + 2\mu\left[P\left(\Delta P + \Gamma Q\right) + Q\left(\Gamma P + \Pi Q\right)\right]$,

$\Omega = \lambda\left(\Theta + \Phi\right)\left(P^2 + Q^2\right) + 2\mu\left[P\left(\Phi P + \Lambda Q\right) + Q\left(\Lambda P + \Theta Q\right)\right]$,

$\Xi = \lambda\left(\Pi + \Delta\right)\left(PR + QS\right) + \mu\left[P\left(\Delta R + \Gamma S\right) + Q\left(\Gamma R + \Pi S\right) + R\left(\Delta P + \Gamma Q\right) + S\left(\Gamma P + \Pi Q\right)\right]$,

$\mathrm{X} = \lambda\left(\Pi + \Delta\right)\left(R^2 + S^2\right) + 2\mu\left[R\left(\Delta R + \Gamma S\right) + S\left(\Gamma R + \Pi S\right)\right]$,

$\Psi = \lambda\left(\Theta + \Phi\right)\left(PR + QS\right) + \mu\left[P\left(\Phi R + \Lambda S\right) + Q\left(\Lambda R + \Theta S\right) + R\left(\Phi P + \Lambda Q\right) + S\left(\Lambda P + \Theta Q\right)\right]$,

$\mathrm{H} = \lambda\left(\Theta + \Phi\right)\left(R^2 + S^2\right) + 2\mu\left[R\left(\Phi R + \Lambda S\right) + S\left(\Lambda R + \Theta S\right)\right]$.

According to (42*f*), one can also obtain

$$ {}_tK^{i'k'} = \frac{1}{{}_tJ} C^{ijkl} F_{k,l}^{k'} F_{i,j}^{i'} = \frac{1}{{}_tJ} F_{i,j}^{i'} F_{k,l}^{k'} \left[ \lambda \delta^{ij} \delta^{kl} + \mu \left( \delta^{ik} \delta^{jl} + \delta^{il} \delta^{jk} \right) \right] = \frac{1}{{}_tJ} \left[ \lambda F_{i,i}^{i'} F_{k,k}^{k'} + \mu \left( F_{k,l}^{i'} F_{k,l}^{k'} + F_{l,k}^{i'} F_{k,l}^{k'} \right) \right] \tag{A11} $$

whose matrix form is

$$ \left[ K^{i'k'} \right] = \begin{bmatrix} C_1 cs^2 + 2C_2 sncs + C_3 sn^2 & -C_2 cs^2 + \left(C_1 - C_3\right) sncs + C_2 sn^2 \\ -C_2 cs^2 + \left(C_1 - C_3\right) sncs + C_2 sn^2 & C_3 cs^2 - 2C_2 sncs + C_1 sn^2 \end{bmatrix} \tag{A12} $$

where $C_1 = \lambda\left(\Pi + \Delta\right)^2 + 2\mu\left(2\Gamma^2 + \Delta^2 + \Pi^2\right)$ , $C_2 = \lambda\left(\Theta + \Phi\right)\left(\Pi + \Delta\right) + 2\mu\left(\Phi\Delta + 2\Lambda\Gamma + \Theta\Pi\right)$ and $C_3 = \lambda\left(\Theta + \Phi\right)^2 + 2\mu\left(2\Lambda^2 + \Phi^2 + \Theta^2\right)$.

## B. The design of an approximate cloak

As Figure 3 shows, the following mappings are adopted to design the wave cloak:

$$ r' = f\left(r\right) \quad \text{and} \quad \theta' = \theta \tag{A13} $$

where $f\left(r\right) = Ar^n + B$ that satisfies $f\left(a\right) = b$, $f\left(c\right) = c$. In addition, the impedance-matching condition on the outer boundary requires $\frac{\mathrm{d}f\left(c\right)}{\mathrm{d}r} = 1$, so that

$$ A = \frac{c - b}{c^n - a^n}, \quad B = \frac{bc^n - a^n c}{c^n - a^n}, \quad Anc^{n-1} = 1. \tag{A14} $$

Noticing $r=\sqrt{x^2+y^2}$ , $\cos\theta=\dfrac{x}{r}$ and $\sin\theta=\dfrac{y}{r}$ in the global Cartesian coordinate system, one obtains

$$x'=r'\cos\theta'=\left(Ar^{n-1}+\frac{B}{r}\right)x \quad\text{and}\quad y'=r'\sin\theta'=\left(Ar^{n-1}+\frac{B}{r}\right)y \tag{A15}$$

Then, the matrix of the deformation gradient can be easily obtained:

$$\left[F_i^{i'}\right]=\begin{bmatrix}\dfrac{Ar^n\left(nx^2+y^2\right)+By^2}{r^3} & \dfrac{\left[Ar^n(n-1)-B\right]xy}{r^3}\\ \dfrac{\left[Ar^n(n-1)-B\right]xy}{r^3} & \dfrac{Ar^n\left(x^2+ny^2\right)+Bx^2}{r^3}\end{bmatrix}=\begin{bmatrix}\dfrac{\Gamma}{r^3} & \dfrac{\Xi}{r^3}\\ \dfrac{\Xi}{r^3} & \dfrac{\mathrm{H}}{r^3}\end{bmatrix} \tag{A16}$$

And ${}_tJ=\det\left(F_i^{i'}\right)=An\left(Ar^n+B\right)r^{n-2}$.

The matrix of metric tensor is

$$\left[{}_tg^{i'j'}\right]=\left[F_i^{i'}F_i^{j'}\right]=\begin{bmatrix}\dfrac{P}{r^4} & \dfrac{R}{r^4}\\ \dfrac{R}{r^4} & \dfrac{Q}{r^4}\end{bmatrix} \tag{A17}$$

where $P=A^2r^{2n}\left(n^2x^2+y^2\right)+\left(2Ar^n+B\right)By^2$ , $Q=A^2r^{2n}\left(x^2+n^2y^2\right)+\left(2Ar^n+B\right)Bx^2$ and

$R=\left[A^2r^{2n}\left(n^2-1\right)-\left(2Ar^n+B\right)B\right]xy$ , with which one can easily obtain ${}_t\rho^{i'j'}=\dfrac{1}{{}_tJ}{}_0\rho\,{}_tg^{i'j'}$ from (42*d*).

According to (A6), the matrix form elastic tensor is:

$$\left[{}_0C^{i'j'k'l'}\right]=\begin{bmatrix}C^{1111} & C^{1122} & C^{1112}\\ & C^{2222} & C^{2212}\\ Sym. & & C^{1212}\end{bmatrix}=\frac{1}{{}_tJ}\begin{bmatrix}\dfrac{(\lambda+2\mu)P^2}{r^8} & \dfrac{\lambda PQ+2\mu R^2}{r^8} & \dfrac{(\lambda+2\mu)PR}{r^8}\\ & \dfrac{(\lambda+2\mu)Q^2}{r^8} & \dfrac{(\lambda+2\mu)QR}{r^8}\\ Sym. & & \dfrac{(\lambda+\mu)R^2+\mu PQ}{r^8}\end{bmatrix} \tag{A18}$$

Then one can use (A8) to obtain the coupling terms

$$\left[D^{1k'l'}\right]=\frac{1}{{}_tJ}\begin{bmatrix}\lambda\dfrac{\Psi Px}{r^7}+2\mu\dfrac{\Gamma(\Omega\Gamma x+\Delta\Xi y)+\Xi(\Sigma\Xi x+\Delta\Gamma y)}{r^{11}} & \lambda\dfrac{\Psi Rx}{r^7}+\mu\dfrac{\Gamma(\Omega\Xi x+\Delta\mathrm{H}y)+\Xi\left[(\Sigma\mathrm{H}+\Omega\Gamma)x+2\Delta\Xi y\right]+\mathrm{H}(\Sigma\Xi x+\Delta\Gamma y)}{r^{11}}\\ Sym. & \lambda\dfrac{\Psi Qx}{r^7}+2\mu\dfrac{\mathrm{H}(\Sigma\mathrm{H}x+\Delta\Xi y)+\Xi(\Omega\Xi x+\Delta\mathrm{H}y)}{r^{11}}\end{bmatrix}$$

$$\left[D^{2k'l'}\right]=\frac{1}{{}_tJ}\begin{bmatrix}\lambda\dfrac{\Psi Py}{r^7}+2\mu\dfrac{\Gamma(\Sigma\Xi x+\Delta\Gamma y)+\Xi(\Sigma\Gamma x+\Pi\Xi y)}{r^{11}} & \lambda\dfrac{\Psi Ry}{r^7}+\mu\dfrac{\Gamma(\Sigma\mathrm{H}x+\Delta\Xi y)+\Xi\left[2\Sigma\Xi x+(\Pi\mathrm{H}+\Delta\Gamma)y\right]+\mathrm{H}(\Sigma\Gamma x+\Pi\Xi y)}{r^{11}}\\ Sym. & \lambda\dfrac{\Psi Qy}{r^7}+2\mu\dfrac{\mathrm{H}(\Sigma\Xi x+\Pi\mathrm{H}y)+\Xi(\Sigma\mathrm{H}x+\Delta\Xi y)}{r^{11}}\end{bmatrix}$$

(A19)

where $\Psi=Ar^n\left(n^2-1\right)-B$ , $\Omega=Ar^n(n-1)\left(nx^2+3y^2\right)-3By^2$ , $\Pi=Ar^n(n-1)\left(3x^2+ny^2\right)-3Bx^2$ ,

$\Delta=Ar^n(n-1)\left[(n-2)x^2+y^2\right]+B\left(2x^2-y^2\right)$ and $\Sigma=Ar^n(n-1)\left[x^2+(n-2)y^2\right]+B\left(-x^2+2y^2\right)$.

According to (A11), one can also obtain

$$\left[K_i^{i'}\right]=\frac{1}{{}_tJ}\begin{bmatrix}\lambda\dfrac{\Psi^2x^2}{r^6}+2\mu\dfrac{\left(\Omega^2+\Sigma^2\right)x^2+2\Delta^2y^2}{r^{10}} & \lambda\dfrac{\Psi^2xy}{r^6}+2\mu\dfrac{(\Omega\Delta+2\Delta\Sigma+\Sigma\Pi)xy}{r^{10}}\\ \lambda\dfrac{\Psi^2xy}{r^6}+2\mu\dfrac{(\Omega\Delta+2\Delta\Sigma+\Sigma\Pi)xy}{r^{10}} & \lambda\dfrac{\Psi^2y^2}{r^6}+2\mu\dfrac{2\Sigma^2x^2+\left(\Delta^2+\Pi^2\right)y^2}{r^{10}}\end{bmatrix} \tag{A20}$$

**References**

1 Leonhardt U. Optical conformal mapping. Science, 2006, 312: 1777-1780

2 Pendry JB, Schurig D, Smith DR. Controlling electromagnetic fields. Science, 2006, 312: 1780-1782

3 Shamonina E, Solymar L. Metamaterials: How the subject started. Metamaterials, 2007, 1: 12-18

4 Greenleaf A, Lassas M, Uhlmann G. On nonuniqueness for Calderon's inverse problem. Math. Res. Lett., 2003, 10: 685-693

5 Greenleaf A, Kurylev Y, Lassas M, Uhlmann G. Invisibility and inverse problems. Bull. Amer. Math. Soc. (N.S.), 2009, 46: 55-97

6 Leonhardt U, Philbin T. Geometry and light. New York: Dover, 2010. 166-170, 173-175

7 Chen H, Chan CT. Acoustic cloaking and transformation acoustics. J. Phys. D: Appl. Phys., 2010, 43: 113001

8 Chen H, Chan CT, Sheng P. Transformation optics and metamaterials. Nature Materials, 2010, 9: 387-396

9 Milton GW, Briane M, Willis JR. On cloaking for elasticity and physical equations with a transformation invariant form. New J. Phys., 2006, 8: 248

10 Ward AJ, Pendry JB. Refraction and geometry in Maxwell's equations. J. Mod. Opt., 1996, 43: 773-793

11 Truesdell C, Noll W. The non-linear field theories of mechanics 3rd Ed. Heideberg: Springer, 2004. 44-47

12 Cummer SA, Schurig D. One path to acoustic cloaking. New J. Phys., 2007, 9: 45

13 Norris AN. Acoustic metafluids. J. Acoust. Soc. Am., 2009, 125: 839-849

14 Norris AN, Shuvalov AL. Elastic cloaking theory. Wave Motion, 2011, 48: 525-538

15 Whitehead AN. The principle of relativity with applications to physical science. Cambridge: Cambridge University Press, 1922. 146-147

16 Willis JR. Variational principles for dynamics problems in inhomogeneous elastic media. Wave Motion, 1981, 3: 1-11

17 Willis JR. Variational principles and operator equations for electromagnetic waves in inhomogeneous media. Wave Motion, 1984, 6: 127-139

18 Willis JR. Dynamics of composites continuum micromechanics. In Suquet P, eds. Continuum micromechanics: CISM Courses and Lectures No. 377. Berlin: Springer-Verlag, 1997. 265-290

19 Willis JR. Effective constitutive relations for waves in composites and metamaterials. Proc. R. Soc. A., 2011, 467: 1865-1879

20 Norris AN, Shuvalov AL, Kutsenko AA. Analytical formulation of three-dimensional dynamic homogenization for periodic elastic systems. Proc. R. Soc. A., 2012, 468: 1629-1651

21 Srivastava A, Nemat-Nasser S. Overall dynamic properties of three-dimensional periodic elastic composites. Proc. R. Soc. A., 2012, 468: 269-287

22 Chen H, Chan CT. Acoustic cloaking in three dimensions using acoustic metamaterials. Appl. Phys. Lett., 2007, 91: 183518

23 Hu J, Chang Z, Hu GK. Approximate method for controlling solid elastic waves by transformation media. Phys. Rev. B., 2011, 84: 201101(R)

24 Cerveny V. Seismic Ray Theory. Cambridge: Cambridge University Press, 2001. 2